\documentclass[%
 reprint,
 amsmath,amssymb,
 aps,
]{revtex4-2}
\usepackage{xcolor}
\usepackage{graphicx}
\usepackage{dcolumn}
\usepackage{bm}
\usepackage{subfigure}
\usepackage{lmodern}
\usepackage[T1]{fontenc}
\usepackage[para]{footmisc}
\usepackage{dblfnote}
\usepackage{dblfloatfix}
\usepackage{placeins}
\usepackage{afterpage}
\makeatother

\begin{document}
\preprint{APS/123-QED}

\title{Optical Memory Optimization Across Rubidium Isotopes and Transitions}

\author{T. Danielov$^1$, I. Puljić$^1$, M. Đujić, D. Aumiler, N. Šantić, T. Ban}
\thanks{Corresponding author: ipuljic@ifs.hr \\ $^1$These authors contributed equally to this work} 
\affiliation{%
 Institute of Physics, Centre for Advanced Laser Techniques, Bijenička cesta 46, 10000 Zagreb, Croatia }





\date{\today}

\begin{abstract}


We investigate optical memory efficiency and storage time across $^{85}\mathrm{Rb}$ and $^{87}\mathrm{Rb}$ isotopes on both the D$_1$ and D$_2$ transitions.
Maximum efficiency of up to $44\%$ was achieved using the D$_1$ line in both isotopes, with up to 1.5 ms storage time. 
These performance levels are enabled by warm vapor rubidium buffer-gas filled cells, large optical depth, and a near-resonant EIT scheme optimized with respect to the one- and two-photon detuning. Our results provide practical guidelines for optimizing the performance of warm rubidium vapor optical memories in simplified experimental configurations. Furthermore, our results provide a foundation for implementing quantum memories in warm alkali vapours with simultaneous optimization of efficiency and storage time.

\end{abstract}

\maketitle


\section{\label{sec:level1}Introduction}


Optical memory enables the on-demand storage and retrieval of light by reversibly mapping photonic states onto long-lived states of matter.
When optical memory is used to store quantum states of light, it is referred to as a quantum memory.
The rapid development of quantum memories was initially driven by proposals for quantum repeaters \cite{Sangouard2011,Wehner2018}, devices which enable the long distance quantum communication.
Beyond this application, quantum memories are also proposed for quantum computing, where they synchronize probabilistic processes and temporarily store intermediate quantum states \cite{Bussieres2013,Liu2023,Heshami2016}, quantum cryptography for enabling the storage and manipulation of quantum keys for secure communication \cite{Mamann2025,mor1999}, and more broadly in quantum information processing \cite{Kaneda2017}. 




A variety of platforms, ranging from doped solid-state systems to cold atomic ensembles, have demonstrated excellent performance in terms of long storage times, high fidelity, and high efficiencies \cite{Ortu2022, Krber2017, Gera2024, Liu2025, Yang2016, Guo2019, Teller2025}. 
However, their implementations are often technically demanding, requiring complex experimental infrastructure and operating procedures.
Because of their operational simplicity and scalability, warm alkali vapors have become a versatile and widely used platform, enabling technological and scientific advances across a broad range of experiments, including light storage.
These vapors, particularly rubidium and cesium, have proven to be highly effective media for optical storage and quantum memory implementations \cite{Ma2022, Wang2022, Esguerra2023, Guo2019,Dideriksen2021}. 
Centimeter-scale vapor glass cells provide high optical depths at moderately low temperatures, while the use of buffer gases and anti-relaxation wall coatings suppresses inelastic wall collisions increasing the atomic coherence times.
A record storage time for classical light on the order of one second, together with a memory efficiency of up to $\sim 10\%$, has been demonstrated in room-temperature cesium vapor \cite{Katz2018}.
On the other hand, a high efficiency of $43\%$ and storage time of around $100~\mu$s has been demonstrated by tedious pulse shaping in warm rubidium vapor \cite{Phillips2008}. 
In optical memories with classical light, optimizing efficiency typically comes at the expense of storage time, and vice versa. A similar trade-off is also observed in quantum memories.
Warm vapor quantum memories in glass cells have achieved efficiencies of up to $\sim 67\%$ using cavity enhanced EIT schemes \cite{Ma2022}, while even higher efficiencies of $\sim 95\%$ have been demonstrated through intelligently shaped input light pulses \cite{Guo2025}.
However, the achieved storage time remains relatively short, e.g. on the order of $\sim 1.6~\mu\mathrm{s}$ in \cite{Ma2022}.
Recently, significant progress in the miniaturization of warm-vapor quantum memories has been achieved through the development of micro-electromechanical systems (MEMS) cells. These cells, which contain alkali vapors, have begun to replace traditional glass cells, with early demonstrations reporting storage efficiencies of a few percent and storage times of several tens of nanoseconds \cite{Mottola2023}. 

While quantum memory experiments have advanced substantially, simultaneously achieving high efficiency and long storage times in warm alkali vapors remains a challenge. 
Here, we addresses this limitation by benchmarking memory efficiency and storage time across rubidium isotopes and optical transitions, using electromagnetically induced transparency (EIT) protocol over a broad experimental parameter space. 
For this purpose, we employed a conventional centimeter-scale vapor cell filled with buffer gas and operated with coupling and probe laser powers typical of diode laser systems, while avoiding complex protocols and extensive optimization.
This study builds upon our previous work \cite{Duji2024} in which we demonstrated that the efficiency of optical storage can be significantly enhanced by operating the EIT protocol in a near-resonant regime. 
In the present study, we perform a detailed investigation of light storage in warm rubidium vapor using such a near-resonant EIT scheme. 
We measured the memory efficiency and storage times on the $5^2S_{1/2} \to 5^2P_{1/2}$ (D$_1$) and $5^2S_{1/2} \to 5^2P_{3/2}$ (D$_2$) transition lines of the $^{85}$Rb and $^{87}$Rb isotopes over a range of experimental parameters, including coupling and probe laser frequencies, probe pulse duration, and rubidium vapor temperature. 
Maximum storage efficiencies of $44\%$ were obtained on the D$_1$ line for both isotopes, and agreed within $1\sigma$. 
The longest memory lifetime of $ 423 \pm 23$ $\mu$s was obtained on the D$_2$ line of the $^{87}$Rb isotope, with up to 1.5 ms storage time.   
To support our experimental findings, we developed a numerical model based on solving the optical Bloch equations for four-level atoms interacting with two laser fields, in accordance with the experimental protocol.

By systematically exploring both isotopes and both the D$_1$ and D$_2$ transitions over a broad parameter space, we identify regimes of moderate yet sufficient efficiency and storage times for applications in the synchronization of optical devices and optical buffering, highlighting the robustness of the near-resonant EIT approach.
In addition, our results optimize a promising platform for the realization of quantum memories with both high efficiency and extended storage time.


The paper is organized as follows. Sec. \ref{sec:setup} describes the experimental setup, and Sec. \ref{sec:theory} outlines the numerical approach based on the optical Bloch equations. In Sec. \ref{sec:results} we present the results and compare the achievable optical memory performance in both rubidium isotopes. Sec. \ref{sec:conclusion} provides a summary.

\section{\label{sec:setup}Experiment}

\begin{widetext}
\begin{figure*}[h!]
\subfigure{\includegraphics[height=7.1cm,width = 0.49\textwidth]{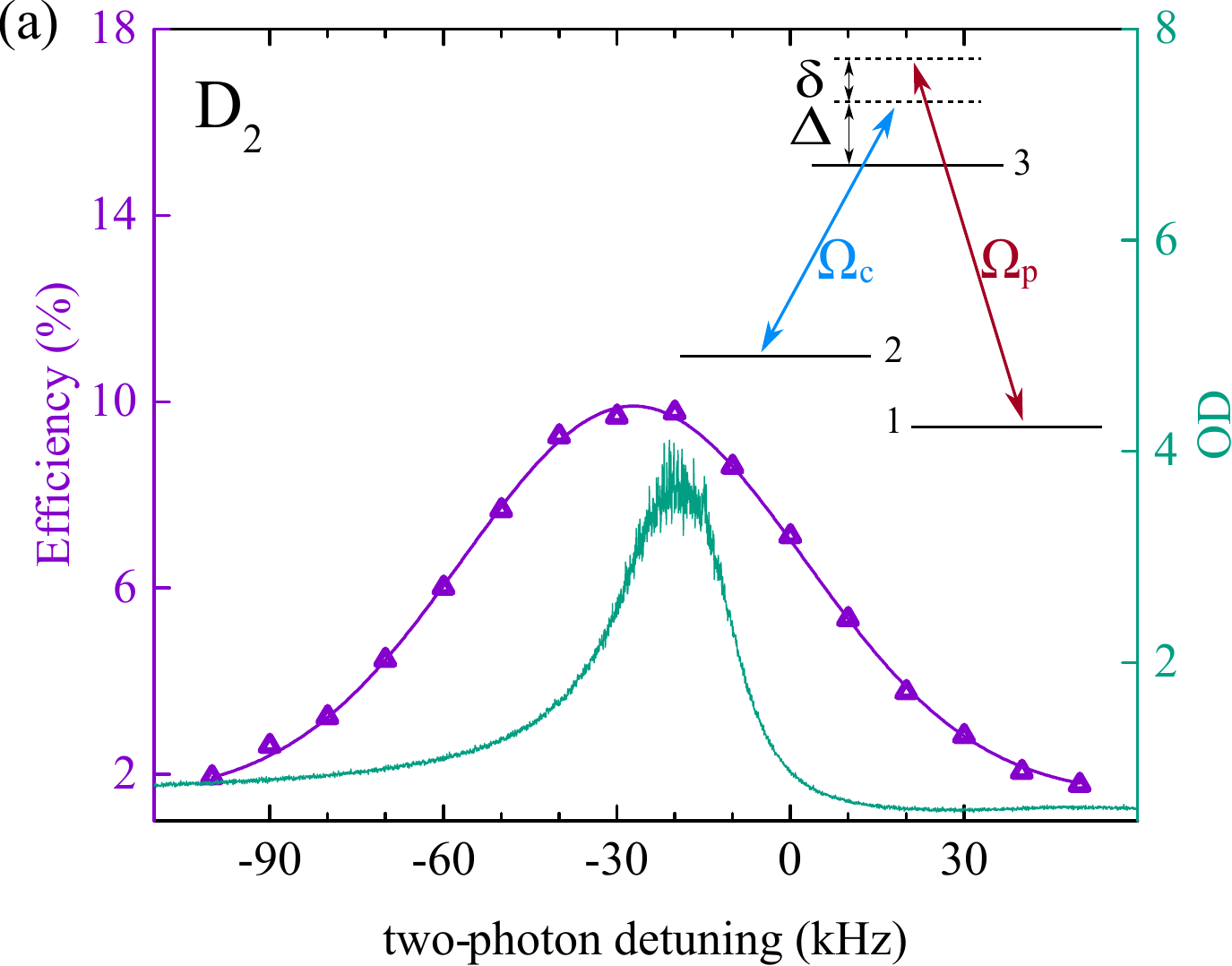}}%
\hspace{0.6cm}
\subfigure{\includegraphics[height=7.1cm,width = 0.4\textwidth]{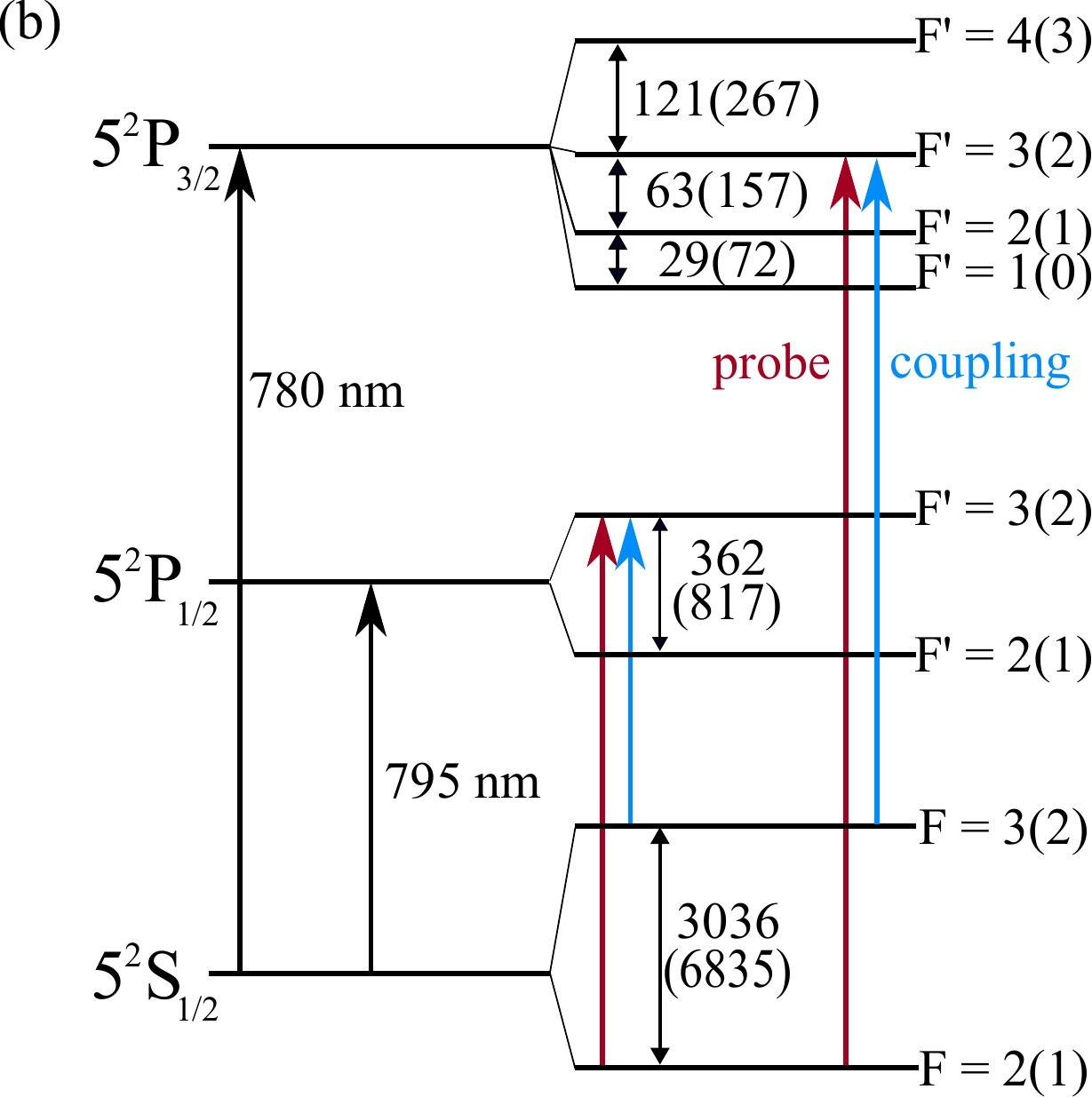}}
\caption[]{\label{fig: levels} (a) Measured memory efficiency $\eta$ (violet) and the corresponding absorption profile (green) on the D$_{2}$ transition of $^{85}$Rb as a function of the two-photon detuning $\delta$, at a one-photon detuning of $\Delta = -700$ MHz and a cell temperature of $45^{\circ}$C. The memory efficiency varies significantly as the two-photon detuning is scanned through resonance. As the one-photon detuning $\Delta$ changes, the EIT lineshape also changes and the optimal two-photon detuning has to be adjusted accordingly. The maximum efficiency for these parameters was observed at $\delta = (-27.2 \pm 0.2)$ kHz.
The inset shows the $\Lambda$-type configuration formed by the hyperfine levels of the D$_1$ and D$_2$ transitions, along with the one-photon and two-photon detunings. 
(b) Hyperfine energy levels of the D$_1$ (795 nm) and D$_2$ (780 nm) transitions in $^{85}$Rb, with the corresponding values for $^{87}$Rb shown in parentheses. During memory measurements on both transitions, the coupling beam was tuned by a one-photon detuning $\Delta$ relative to the $F = 3(2) \rightarrow F'= 3 (2)$ transition, while the probe beam was tuned by $\Delta + \delta$ relative to the $F = 2(1) \rightarrow F'= 3 (2)$ transition. The arrows indicate the probe (red) and coupling (blue) transitions, and the hyperfine splittings of the levels are given in MHz.  }
\end{figure*}
\end{widetext}

In our experiment, we measured the efficiency and storage time of an optical memory in warm vapors of $^{85}$Rb and $^{87}$Rb using a near-resonant EIT protocol. 
The measurements were performed on the D$_1$ ($5^{2}S_{1/2} \rightarrow 5^{2}P_{1/2}$) and D$_2$ ($5^{2}S_{1/2} \rightarrow 5^{2}P_{3/2}$) transitions for each isotope. 
A $\Lambda$-type configuration is employed, formed by two hyperfine levels of the ground $5^{2}S_{1/2}$ state and one hyperfine level of the excited state, $5^{2}P_{1/2}$ or $5^{2}P_{3/2}$, depending on whether the D$_1$ or D$_2$ transition is used, respectively.
In a $\Lambda$ scheme two detunings must be precisely controlled, as shown in Fig. \ref{fig: levels}(a): the one-photon detuning $\Delta$ and the two-photon detuning $\delta$.
The one-photon detuning is defined as $\Delta= \omega_c - \omega_{23}$, where $\omega_c$ is the frequency of the coupling laser and $\omega_{23}$ is the transition frequency between levels $2$ and $3$.
The two-photon detuning $\delta$ quantifies the deviation from the two-photon resonance condition, and 
is defined as $\delta = (\omega_p - \omega_c) - \omega_{12}$, where $\omega_p$ is the frequency of the probe laser and $\omega_{12}$ is the transition frequency between levels $1$ and $2$.

When measuring the absorption of a weak probe field in the presence of a strong coupling laser near $\Delta \approx 0$, EIT is observed \cite{DeRose2023,Finkelstein2023}, whereby an otherwise absorbing atomic medium becomes transparent to the probe light. 
This effect is accompanied by a steep dispersion of the refractive index, leading to a significant reduction in the group velocity of light propagating through the medium.
In buffer-gas filled cells, EIT exhibits a symmetric lineshape around $\Delta = 0$, which can be described by a Lorentzian profile.
As $|\Delta|$ increases from zero, the shape of the EIT resonance line evolves from a symmetric transparency peak to a dispersion-like feature that gradually transforms into a narrow absorption resonance corresponding to a two-photon Raman transition \cite{Duji2024, Mikhailov2004}. 
In a three-level scheme, $\delta$ for which this two-photon absorption resonance is achieved is proportional to the coupling laser intensity and inversely proportional to $\Delta$. For large enough $\Delta$, the width of this two-photon resonance is limited by the ground level coherence decay rate $\gamma_{12}$ \cite{Mikhailov2004}. 
This behavior can lead to improved memory performance in far-detuned EIT protocols compared to the resonant case, as demonstrated in our previous work \cite{Duji2024} and further investigated in detail in this study.

In all our measurements, for a given $|\Delta|$, the memory efficiency was optimized with respect to the two-photon detuning $\delta$. 
An example of such a measurement is shown in Fig. \ref{fig: levels}(a), where we present the probe absorption and memory efficiency on the D$_2$ transition of $^{85}$Rb as a function of the two-photon detuning $\delta$ for $\Delta = -700$~MHz. 
Here, $\Delta$ denotes the detuning of the coupling laser from the $F=3 \rightarrow F' = 3$ transition, while the probe laser is scanned over a range of $\delta$ around the $F=2 \rightarrow F' = 3$ + $\Delta$ frequency. 
$F$ denotes the hyperfine level of the ground $5^{2}S_{1/2}$ state, and $F'$ denotes the hyperfine level of the excited $5^{2}P_{3/2}$ state in $^{85}\mathrm{Rb}$. 
A detailed energy-level scheme of the hyperfine structure for the D$_1$ ($5^{2}S_{1/2} \rightarrow 5^{2}P_{1/2}$) and D$_2$ ($5^{2}S_{1/2} \rightarrow 5^{2}P_{3/2}$) transitions in $^{85}\mathrm{Rb}$ and $^{87}\mathrm{Rb}$ is shown in Fig.~\ref{fig: levels}(b) \cite{Steck2008Rubidium85,Steck2003Rubidium87}.
The measured efficiency as a function of $\delta$ is well described by a Gaussian curve with a width approximately three times larger than that of the two-photon resonance. 
The maximum efficiency is observed for $\delta= (-27.2 \pm 0.2)$ kHz and it is shifted toward the red side with respect to the peak of the two-photon resonance.

Our experimental setup follows a design similar to that of our previous work \cite{Duji2024}, and is only briefly described here. 
A frequency stabilized titanium-sapphire laser is split into two orthogonally $\pi$-polarized beams - a high-intensity coupling beam and a low-intensity probe beam. 
The probe beam is sent through a fiber-coupled electro-optical modulator (EOM) and modulated at a frequency equal to the hyperfine splitting of the $^{85}\mathrm{Rb}$ ($^{87}\mathrm{Rb}$) $5^2S_{1/2}$ ground-state, generating frequency sidebands offset from the carrier by that splitting. The output beam from the EOM is transmitted through an optical filtering cavity that rejects all frequency components except the blue sideband, which serves as the probe input for optical memory realization.  

Acousto-optical modulators (AOMs) in a double-pass configuration are used for both beams, enabling fast switch-off times, independent frequency tuning, and precise control of the one-photon detuning $\Delta$ and the two-photon detuning $\delta$.
Polarization optics were used to combine the coupling and probe beams before entering the vapor cell and to separate the strong coupling field from the weak probe field after propagation through the atomic medium. 
The $1/e^2$ diameter of the coupling beam was set to $d_C = 2.5$ cm, while the probe beam diameter was $d_P = 1.1$ cm. 
For the $^{85}$Rb D$_2$ transition, the optical powers were $P_C = 22$ mW and $P_P = 4.4~\mu$W, corresponding to intensities of $I_C = 4.5$ mW/cm$^2$ and $I_P = 4.63~\mu$W/cm$^2$, and resulting in Rabi frequencies \cite{Steck2008Rubidium85} of $\Omega_C = 2\pi \times 5.7$ MHz and $\Omega_P = 2\pi \times 0.2$ MHz for the coupling and probe beams, respectively.
For measurements on the D$_1$ line of $^{85}\mathrm{Rb}$, as well as on the $^{87}\mathrm{Rb}$ isotope, the coupling and probe laser powers were adjusted to maintain the corresponding Rabi frequencies at the values given above.
After propagating through the cell, the probe light is detected using an avalanche photodiode (Thorlabs, APD410A/M).

Two different cylindrical quartz glass cells were used, one containing pure $^{85}$Rb and the other containing pure $^{87}$Rb. 
Both cells have a length of 7.5 cm and a diameter of 2.5 cm, and are filled with 5 Torr of neon buffer gas, with paraffin anti-relaxation coating on the walls. 
The cell is placed inside a three-layered $\mu$-metal shield and heated externally using copper heat pipes. 
The temperature is stabilized using a PID loop, allowing precise control of the vapor temperature.
The optical depth is determined from the absorption spectra using the Beer-Lambert law \cite{demtroder2010}
by recording the transmission of the probe laser while scanning its frequency across the desired transition at a given temperature. 
The maximum operating temperature of $60^{\circ}$C is limited by the anti-relaxation coating, which degrades at higher temperatures.

The measurement protocol follows the same procedure as in our previous work \cite{Duji2024}. First, the strong coupling beam is turned on. After a short delay, an exponentially rising probe pulse with a full-width half-maximum (FWHM) of $6~\mu$s is applied. This duration was optimized for maximum memory efficiency, as discussed in \cite{Duji2024}. The probe and coupling beams are switched off simultaneously and remain off for a duration referred to as the storage time. After this delay, the coupling beam is turned on again, triggering the retrieval of the stored probe pulse. 
The precise timing of the sequence, as well as the shaping of the exponential probe and square coupling pulses, are controlled by the AOMs driven by a common two-channel RF synthesizer (Moglabs, XRF421), with a timing resolution of 16 ns. 
For each set of experimental conditions, the protocol was repeated 5000 times to obtain averaged time-resolved photodiode traces. 
The memory efficiency is then calculated as the ratio of the retrieved pulse energy to that of the incident probe pulse, and the memory lifetime is determined from the storage time at which the efficiency decays to $1/e$.

\section{Numerical model} \label{sec:theory}
Our numerical approach to the optical memory protocol on the D$_1$ transition of both isotopes is based on solving the optical Bloch equations (OBEs) of a four-level atomic system interacting with two laser fields. One field, $\mathcal{E}_c$, corresponds to the high intensity coupling beam, while the second field, $\mathcal{E}_p$, corresponds to the low intensity probe beam. We number the energy states from $1$ to $4$, where the lowest (highest) number corresponds to the lowest (highest) energy level, see Fig. \ref{fig: levels}. The time evolution of the density matrix elements, $\rho_{nm}$, is expressed by a set of sixteen OBEs:  
\begin{align}
    \dfrac{\partial\rho_{nm}}{\partial t} &= -\dfrac{i}{\hbar}\left[\hat{H},\hat{\rho}\right]_{nm}-\gamma_{nm}\rho_{nm}, \quad n\neq m \label{eq1} \\
    \dfrac{\partial\rho_{nn}}{\partial t} &= -\dfrac{i}{\hbar}\left[\hat{H},\hat{\rho}\right]_{nn} - \sum_{m<n}\Gamma_{mn}\rho_{nn} \label{eq2}\\
    & \phantom{=}+\sum_{m>n}\Gamma_{nm}\rho_{mm} \notag
\end{align}

where $n,m=\{1,2,3,4\}$, $\Gamma_{nm}$ are the population decay rates from level $m$ to level $n$ and $\gamma_{nm} = \frac{1}{2}(\Gamma_n + \Gamma_m)+\gamma^{coll}_{nm}$ are coherence damping factors of $\rho_{nm}$ density matrix elements. Here, $\Gamma_1 = \Gamma_2=0$ and $\Gamma_3 = \Gamma_4 = 2 \pi \times 5.75$ MHz for both isotopes \cite{Steck2008Rubidium85, Steck2003Rubidium87}, are the total population decay rates. The spin exchange collisions $\gamma_{12}^{coll} = \gamma_{21}^{coll} = \gamma_{34}^{coll} = \gamma_{43}^{coll}$  were taken to be 0.5 kHz, estimated from the average of the measured lifetimes on D$_1$ and D$_2$ transitions of both isotopes. The $\gamma_{nm}^{coll}$ of optical transitions, i.e. $\gamma_{13}^{coll} = \gamma_{23}^{coll} = \gamma_{14}^{coll} = \gamma_{24}^{coll}$, include Doppler broadening and pressure broadening \cite{Duji2024}. 
Doppler broadening for anti-relaxation coated neon-filled cell at a temperature of $60^{\circ}$C was taken to be $2\pi \times 535$ MHz \cite{demtroder2010} and pressure broadening for a cell filled with 5 Torr of Ne gas was $2\pi \times 25$ MHz \cite{DeRose2023}.
The total Hamiltonian of the system $\hat{H}$ is a sum of the diagonalized free atom Hamiltonian $\hat{H}_0$ and the interaction Hamiltonian $\hat{H}_{\text{int}}$: $\hat{H}=\hat{H}_0+\hat{H}_{\text{int}}$. The interaction of the atom with the laser fields can be expressed in the dipole approximation as $\hat{H}_{\text{int}}=-\sum_{n}(\mu_{1n}\mathcal{E}_p(t)+\mu_{2n}\mathcal{E}_c(t)$ + H.c.), where $n=3,4$. The coefficients $\mu_{mn}$ represent the transition dipole moments of corresponding transitions. Transition dipole moments, hyperfine level splittings, and population decay rates are taken from \cite{Steck2008Rubidium85, Steck2003Rubidium87}, depending on the rubidium isotope.

To solve the optical Bloch equations, we write the probe and coupling field as $\mathcal{E}_p(t)=E_pe^{-i\omega_pt}$ and $\mathcal{E}_c(t)=E_ce^{-i\omega_ct}$, where probe field amplitudes $E_p$, corresponded to  8.5 V/m for $^{85}$Rb and 8.2 V/m for $^{87}$Rb, matching the experimental values. Amplitudes of the pump fields were set to 200 V/m and 180 V/m, for $^{85}$Rb and $^{87}$Rb respectively, yielding the best match to the experimentally measured efficiency curves. Frequencies $\omega_p$ and $\omega_c$ were rewritten as $\omega_p=\omega_{14}+\Delta+\delta$ and $\omega_c=\omega_{24}+\Delta$, where $\Delta$ is one-photon, and $\delta$ is two-photon detuning. The two-photon detunings $\delta$ were fixed to the corresponding measured values, i.e. negative for red one-photon detunings, and positive for positive one-photon detuings.

\begin{figure}[h!]
\includegraphics[width = 0.85\columnwidth]{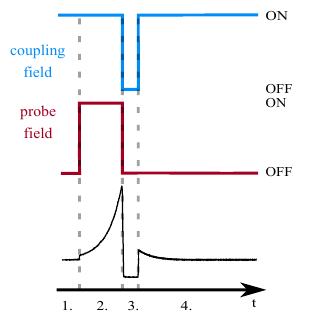}
\caption{\label{fig: protocol} Up: The time domain of our protocol is divided into four parts: the coupling field (blue) is on for duration of $100~\mu$s, then the probe field (red) is turned on for $6~\mu$s after which both fields are switched off and stay off for a duration of $5~\mu$s. In the last part of the protocol, the coupling field is turned on again for $100~\mu$s and the transmission of the probe electric field is calculated. Down: Probe signal in different domains of the experimental protocol, 
measured on a photodiode. 
}
\end{figure}

To calculate the efficiency of our optical memory, we split the time domain of the protocol into four parts, as shown in Fig. \ref{fig: protocol}, following the experimental protocol
. In the first domain, the atoms interact only with the electric field of the coupling laser for a duration of $100~\mu$s. In the second domain, in addition to the coupling field, the electric field of the probe laser is switched to a finite value for a duration of $6~\mu$s, corresponding to the FWHM duration of the probe pulse in our experiment. In the third domain, both the coupling and the probe fields are simultaneously switched off and remain off for a duration known as the storage time, set to the experimental value of $5~\mu$s. The fourth domain of the protocol is the retrieval of the stored excitation, where only coupling field is turned on to its initial value for $100~\mu$s.

For a given one-photon detuning $\Delta$ we solve the optical Bloch equations in each time domain. 
The obtained density matrix elements from each domain are then used as initial conditions for the following one. The solutions of the off-diagonal matrix elements from the fourth domain are used to calculate the electric field of the retrieved probe pulse, with the dominant term originating from $\rho_{14}$. Memory efficiency is defined as the ratio of time-integrated squared electric field
of the retrieved probe pulse to that of the input probe
pulse. 

\begin{figure}[h!]
\includegraphics[width = \columnwidth]{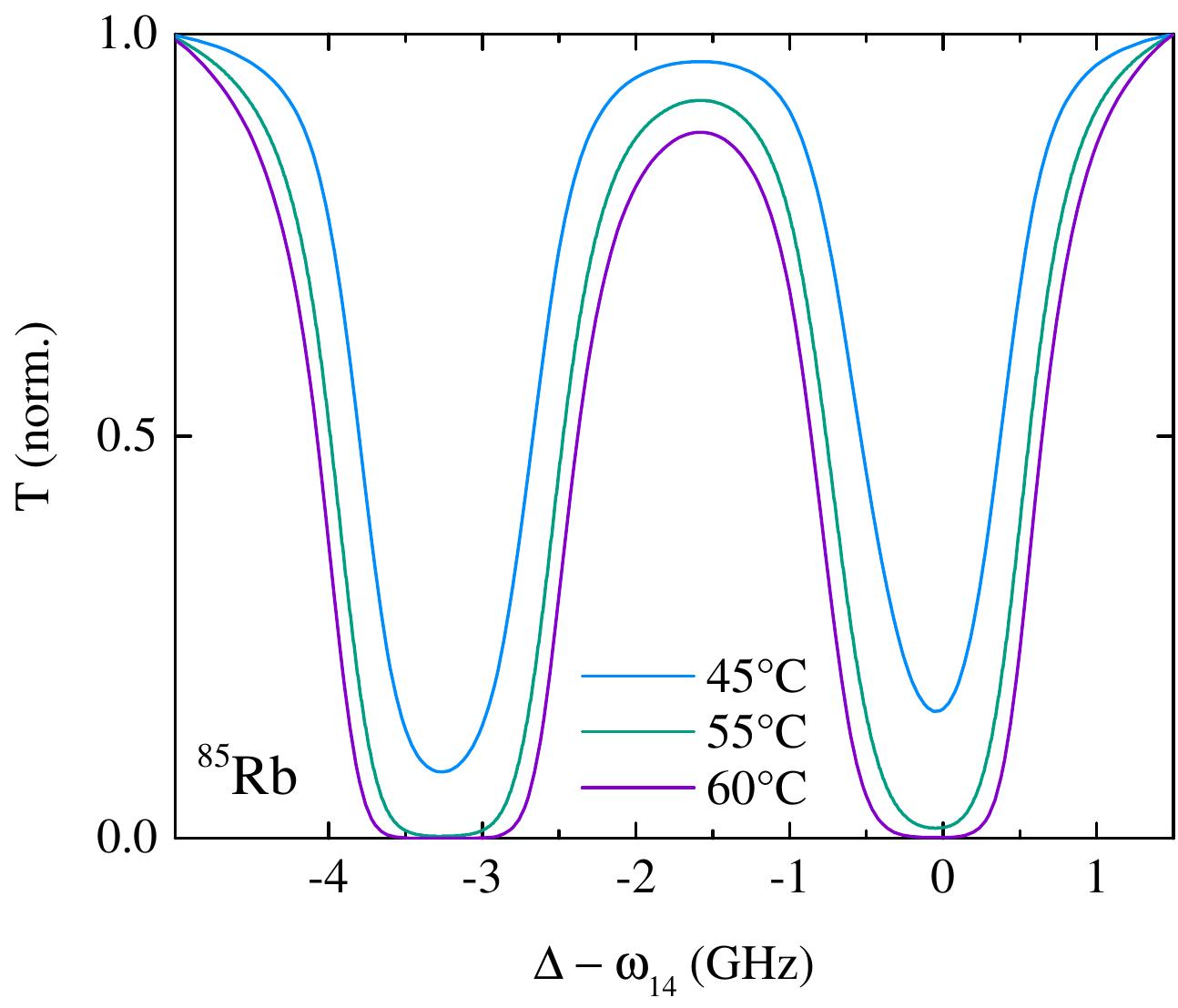}
\caption{\label{fig: transmission} Calculated normalized transmission of the weak laser field in dependence of one-photon detuning $\Delta$ for D$_1$ transition of $^{85}$Rb at temperatures of $45^{\circ}$C (blue), $55^{\circ}$C (green) and $60^{\circ}$C (violet). 
The coupling laser intensity was corrected to account for absorption within the medium, for each value of $\Delta$.
}
\end{figure}

For each value of $\Delta$, the coupling laser intensity was corrected  to account for absorption within the medium. To determine this correction, the OBEs for a four level atomic system interacting with a single laser field were solved, and the resulting off-diagonal matrix elements were used to determine the imaginary part of the refractive index. This enabled the calculation of the attenuation of the laser intensity during propagation through the atomic medium, from which the effective coupling laser electric field used in solving the optical Bloch equations for the memory
protocol was determined. 
The calculated normalized transmission spectra for D$_1$ transition of $^{85}$Rb at temperatures of $45^{\circ}$C, $55^{\circ}$C  and $60^{\circ}$C are shown in Fig. \ref{fig: transmission}, corresponding to optical depths of 2.6, 6.2, 9.5, respectively. The transmission spectra was measured on the $D_1$ line of $^{85}$Rb at room temperature and showed very good agreement with numerically obtained curve. For higher temperatures the optical depth was therefore calculated from corresponding concentration of atoms.

\section{\label{sec:results}Results}

In this section we present measurements of the memory efficiency and lifetime in warm vapors of $^{85}$Rb and $^{87}$Rb. 
For each isotope, $\Lambda$-type configuration formed by hyperfine levels of the D$_1$ and D$_2$ transitions is utilized.
For $^{85}\mathrm{Rb}$, in both the D$_1$ and D$_2$ cases, the coupling laser was tuned to the $F=3 \rightarrow F' = 3$ transition and the probe laser to $F=2 \rightarrow F' = 3$, when $\Delta = 0$ (see Fig. \ref{fig: levels}(b)). 
Similarly, for $^{87}\mathrm{Rb}$, the coupling and probe lasers were tuned to the $F=2 \rightarrow F' = 2$ and $F=1 \rightarrow F' = 2$ transitions, respectively (see Fig. \ref{fig: levels}(b)).

Measurements are performed in isotope-enriched vapors of $^{85}\mathrm{Rb}$ and $^{87}\mathrm{Rb}$, each contained in a separate cell at a temperature of $60^{\circ}\mathrm{C}$. 
In case of $^{85}\mathrm{Rb}$ D$_1$ transition, this temperature corresponds to an optical depth of OD = 9.5.

\begin{widetext}
\begin{figure*}[h!]
\includegraphics[height = 7.1cm,width = 0.99\textwidth]{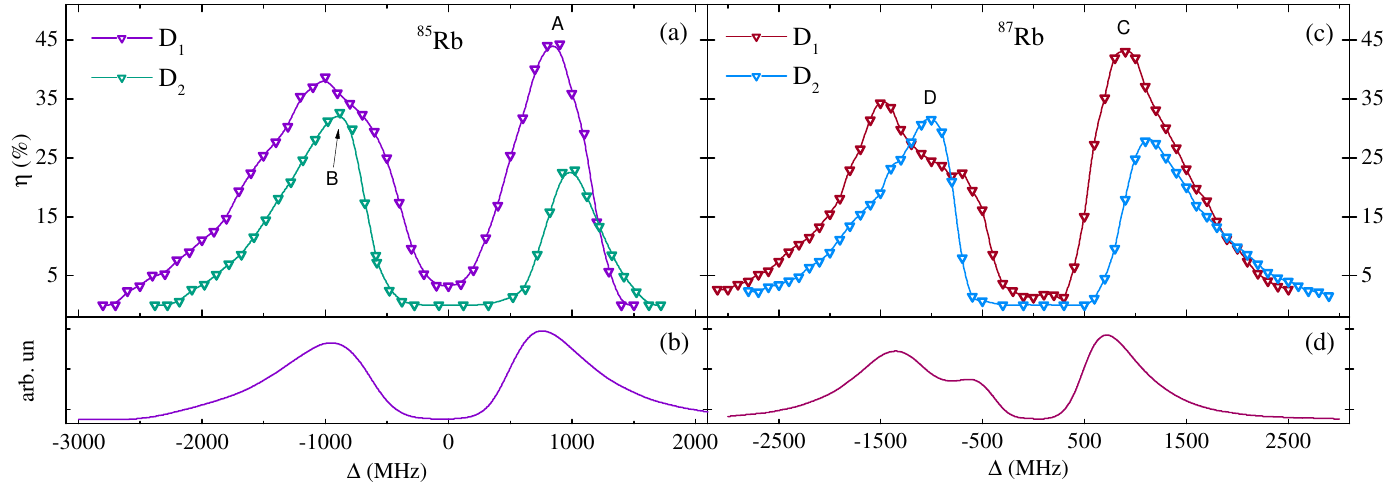}
\caption{\label{fig:one_photon_scans 85rb}  Measured memory efficiencies $\eta$  operating on the D$_1$ (violet, red) and D$_2$ (green, blue) atomic transitions of $^{85}$Rb (a) and $^{87}$Rb (c), as a function of the one-photon detuning $\Delta$. Measurements were performed at the cell temperature of $60^{\circ}$C, corresponding to the optical depth of OD = 9.5. Powers of the coupling and the probe beam were adjusted so that the Rabi frequencies corresponded to $\Omega_{C} = 2 \pi \times 5.7$ MHz and $\Omega_{P} = 2 \pi \times 0.2$ MHz, respectively. The two-photon detuning was optimized for each data point, and storage time was set to 5 $\mu$s. Corresponding numerical model results on the D$_1$ transition for $^{85}$Rb (b) and $^{87}$Rb (d), respectively. }
\end{figure*}
\end{widetext}

\textbf{Memory efficiency.} In Fig. \ref{fig:one_photon_scans 85rb}(a) we show the measured memory efficiencies $\eta$ for the $^{85}$Rb D$_1$ (violet) and D$_2$ (green) transitions as a function of the one-photon detuning $\Delta$.
For each $\Delta$, the memory efficiency shown was optimized with respect to the two-photon detuning $\delta$ and it is measured for the storage time of $5$ $\mu$s.
For both transitions, the measured memory efficiency as a function of $\Delta$ exhibits a double-peak structure resulting from two effects that significantly influence the storage efficiency: absorption of the coupling beam and the number of atoms participating in the storage process. 
For $\Delta \approx 0$, due to the large OD, the coupling laser is strongly absorbed during its propagation through the cell. 
This implies that the atoms effectively experience a low average coupling intensity, since the laser intensity decays exponentially along the cell according to the Beer–Lambert law, which in turn leads to a significant reduction in memory efficiency.
As $|\Delta|$ increases from zero, the absorption of the coupling laser decreases and the intensity experienced by the atoms increases, leading to a higher memory efficiency.
However, varying $|\Delta|$ also 
changes velocity class addressed within the Maxwell-Boltzman distribution, such that larger $|\Delta|$ leads to a reduced number of atoms participating in the interaction. 
The one-photon detuning $\Delta$ at which the maximum efficiency is observed therefore represents a "sweet spot", where the effective coupling intensity experienced by the atoms and the number of atoms participating in the interaction are optimally balanced for the storage process.

However, from inspection of Fig. \ref{fig:one_photon_scans 85rb}(a), it is evident that the efficiency curve around $\Delta = 0$ is not symmetric, as would be expected from the arguments presented above. 
For the D$_1$ transition of both isotopes, the memory efficiency is higher for $\Delta > 0$, i.e., when the coupling laser is tuned to the blue side of the D$_1$ $F=3 \rightarrow F' = 3$ transition. 
In contrast, for the D$_2$ line, higher efficiency is observed for $\Delta < 0$, corresponding to the red-detuned side of the D$_2$ $F=3 \rightarrow F' = 3$ transition. 
This asymmetry arises from the hyperfine structure of the $5^{2}P_{1/2}$ and $5^{2}P_{3/2}$ manifolds, which constitute the excited states of the D$_1$ and D$_2$ transitions, respectively.

In the case of the $5^{2}P_{1/2}$ state (D$_1$ transition), a nearby hyperfine level $F' = 2$ which lies at lower energy than the $F' = 3$ level, from which $\Delta$ is defined, exists, as shown in Fig. \ref{fig: levels}(b). 
The energy separation between these two excited hyperfine levels is approximately 362~MHz, which is smaller than the Doppler width of about 535 MHz for $^{85}\mathrm{Rb}$ atoms at a temperature of $60^\circ\mathrm{C}$ \cite{demtroder2010}. 
Consequently, for a given $|\Delta|$, the coupling field experiences increased absorption for $\Delta < 0$, which reduces the memory efficiency.
In contrast, for the $5^{2}P_{3/2}$ state (D$_2$ transition), the $F' = 4$ hyperfine level has approximately 121~MHz higher frequency than the $F' = 3$ level, as shown in Fig. \ref{fig: levels}(b), and the transition dipole moment for $F = 3 \rightarrow F' = 4$ transition is 
larger than for the $F = 3 \rightarrow F' = 3$ transition.
Although below the $F' = 3$ level there are also the $F' = 1$ and $F' = 2$ hyperfine levels, the $F = 3 \rightarrow F' = 1$ transition is dipole-forbidden, while the $F' = 2$ level lies approximately 63~MHz below the $F' = 3$ level and has a smaller transition dipole moment \cite{Steck2008Rubidium85}. 
As a result, the absorption of the coupling laser becomes asymmetric with respect to the $F' = 3$ level, leading to stronger absorption for $\Delta > 0$ and consequently reduced memory efficiency in that detuning range.

Moreover, it is worth noting that the presence of additional hyperfine levels in the excited state allows for the simultaneous realization of two $\Lambda$ schemes, whose relative contributions to the memory efficiency depend on $\Delta$. 
For example, in the case of the D$_1$ transition, both $\Lambda$ schemes contribute to the efficiency for $\Delta < 0$, whereas for $\Delta > 0$ the scheme associated with the $F' = 3$ level becomes dominant.
This results in a broader memory efficiency profile for $\Delta < 0$ compared to the case of $\Delta > 0$.

And finally, by comparing the spectral profiles of the measured memory efficiency for the D$_1$ and D$_2$ lines, we observe that the dip around $\Delta = 0$ is broader for the D$_2$ transition than for D$_1$. 
This can be attributed to the larger transition dipole matrix element of the D$_2$ line \cite{Steck2008Rubidium85}, which leads to stronger absorption of the coupling field. 
Consequently, for the same detuning $\Delta$, the coupling beam is more strongly absorbed on the D$_2$ transition, resulting in a reduced memory efficiency compared to D$_1$.
This also explains the higher memory efficiency observed for the D$_1$ transition over the entire range of $\Delta$.

The highest memory efficiency for the $^{85}\mathrm{Rb}$ isotope, $\eta(^{85}\mathrm{Rb}) = (44 \pm 3)\%$, is obtained on the D$_1$ transition at a one-photon detuning of $\Delta = 900$~MHz and a two-photon detuning of $\delta = 30$~kHz.
The quoted uncertainty in memory efficiency was determined from six independent measurements performed under the same experimental conditions.

In fig. \ref{fig:one_photon_scans 85rb} (c) we show the measured memory efficiencies $\eta$ for the $^{87}$Rb D$_1$ (red) and D$_2$ (blue) transitions as a function of the one-photon detuning $\Delta$.
For each $\Delta$, the memory efficiency is optimized with respect to the two-photon detuning $\delta$, while the Rabi frequencies of the coupling and probe beams, as well as the storage time, are kept the same as in the $^{85}\mathrm{Rb}$ case.
The dependence of the memory efficiency on $\Delta$ is very similar to that observed for the $^{85}\mathrm{Rb}$ isotope for both the D$_1$ and D$_2$ transitions.
In the case of the D$_1$ line, the contribution of the two $\Lambda$ schemes, i.e. $F = 2 \rightarrow F' = 2$ and $F = 2 \rightarrow F' = 1$, to the memory efficiency is clearly visible. The energy separation between the $F' = 1$ and $F' = 2$ levels is approximately 817~MHz~\cite{Steck2003Rubidium87}, which exceeds the Doppler linewidth and consequently results in two distinct peaks in the memory efficiency being resolved for $\Delta < 0$.

The highest memory efficiency for the $^{87}\mathrm{Rb}$ isotope, $\eta(^{87}\mathrm{Rb}) = (43 \pm 3)\%$, is obtained on the D$_1$ transition at a one-photon detuning of $\Delta = 900$~MHz and a two-photon detuning of $\delta = 10$~kHz.

The results of the numerical model, obtained using parameters corresponding to the experimental conditions (see Sec. \ref{sec:theory}), show good qualitative agreement with the experiment for the $D_1$ transition of both isotopes, as illustrated in Figs.~\ref{fig:one_photon_scans 85rb} (b) and (d). 
The model provided the best agreement with the experimental data for a cell length of 7.5 cm and a rubidium vapor temperature of $55^{\circ}$C for both isotopes. 

\begin{figure}[b]
\includegraphics[width = \columnwidth]{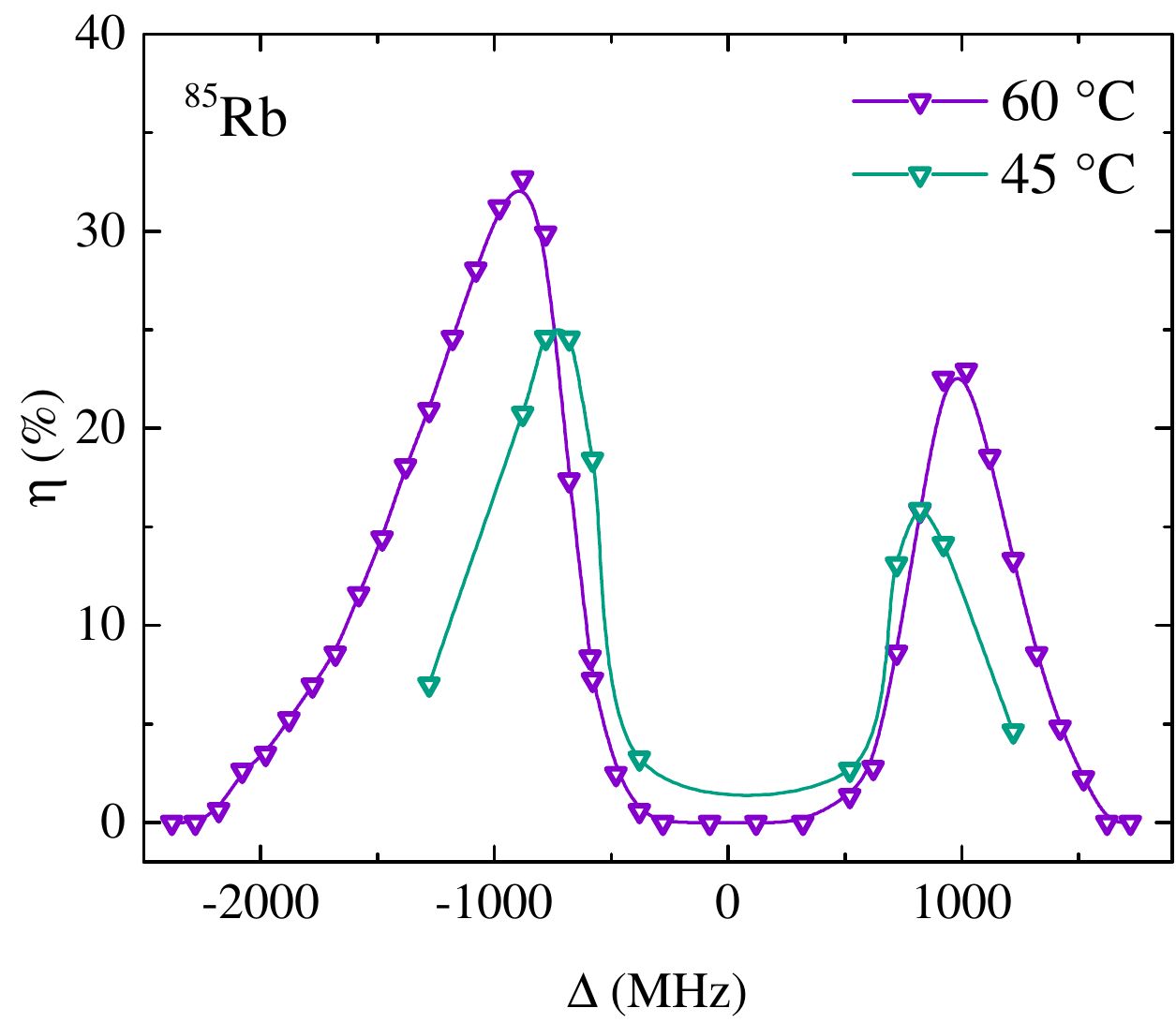}
\caption{\label{fig:temperature}  Measured memory efficiencies $\eta$ operating on the D$_2$ atomic transition of $^{85}$Rb, as a function of the one-photon detuning $\Delta$, at the cell temperature of $60^{\circ}$C (violet) and $45^{\circ}$C (green). This corresponds to the optical depths of OD = 9.5 and OD = 2.6, respectively. Power of the coupling and the probe beams were adjusted so that the Rabi frequencies corresponded to $\Omega_{C} = 2 \pi \times 5.7$ MHz and $\Omega_{P} = 2 \pi \times 0.2$ MHz, respectively. The two-photon detuning was optimized for each data point, and storage time was set to 5 $\mu$s.}
\end{figure}

In Fig. \ref{fig:temperature}, we present the measured memory efficiency $\eta$ for the $^{85}\mathrm{Rb}$ D$_2$ transition as a function of the one-photon detuning $\Delta$, for two cell temperatures: $60^{\circ}\mathrm{C}$ (violet) and $45^{\circ}\mathrm{C}$ (green), corresponding to optical depths of OD = 9.5 and OD = 2.6, respectively. 
All other experimental parameters are kept the same to those used in the previously shown measurements. 
For each data point, the two-photon detuning is optimized, while the storage time is fixed at $5~\mu\mathrm{s}$.
From Fig. \ref{fig:temperature}, it is evident that the efficiency peaks are higher and shift toward larger values of $|\Delta|$ at higher temperature. 
This behavior arises from the interplay between absorption of the coupling beam, leading to a reduction in its intensity, and the increase in number of atoms.
For a given one-photon detuning $\Delta$, the absorption of the coupling laser increases with the increase in the number of atoms. 
For example, at $\Delta = 1000\,\mathrm{MHz}$, the transmitted intensity of the coupling laser is approximately 1.5 times higher at $45\,^{\circ}\mathrm{C}$ than at $60\,^{\circ}\mathrm{C}$, as determined from calculated transmission curves, see Sec. \ref{sec:theory}.
The impact of the coupling laser absorption on memory efficiency is clearly visible in the vicinity of $\Delta = 0$. 
At higher temperatures, increased absorption of the coupling laser leads to reduced efficiency at $60\,^{\circ}\mathrm{C}$ compared to $45\,^{\circ}\mathrm{C}$, while the width of the observed dip in the efficiency curve increases.
The higher peak efficiencies observed at $60\,^{\circ}\mathrm{C}$ are attributed to the increased number of atoms participating in the interaction at elevated vapor temperatures. 
The number of atoms is 
defined by the vapor temperature, and in our case is 3.7 times higher at $60\,^{\circ}\mathrm{C}$ than at $45\,^{\circ}\mathrm{C}$.
This also contributes to the increased efficiency in the outer wings of the peaks.

In our experiment, the maximum achievable temperature of $60^{\circ}\mathrm{C}$ was limited by the paraffin coating of the cell. 
However, our results suggest that higher memory efficiencies could be achieved using buffer-gas cells without paraffin coating, allowing operation at higher temperatures and thus increasing the number of atoms, which in turn increases memory efficiency. 
We also note that in this regime, the efficiency peak is expected to shift toward larger one-photon detuning values $|\Delta$|, where, again, an optimal balance between the effective coupling intensity and the number of atoms participating in the interaction will be reached.



\begin{widetext}
\begin{figure*}[h!]
\includegraphics[height = 7.1cm,width = 0.99\textwidth]{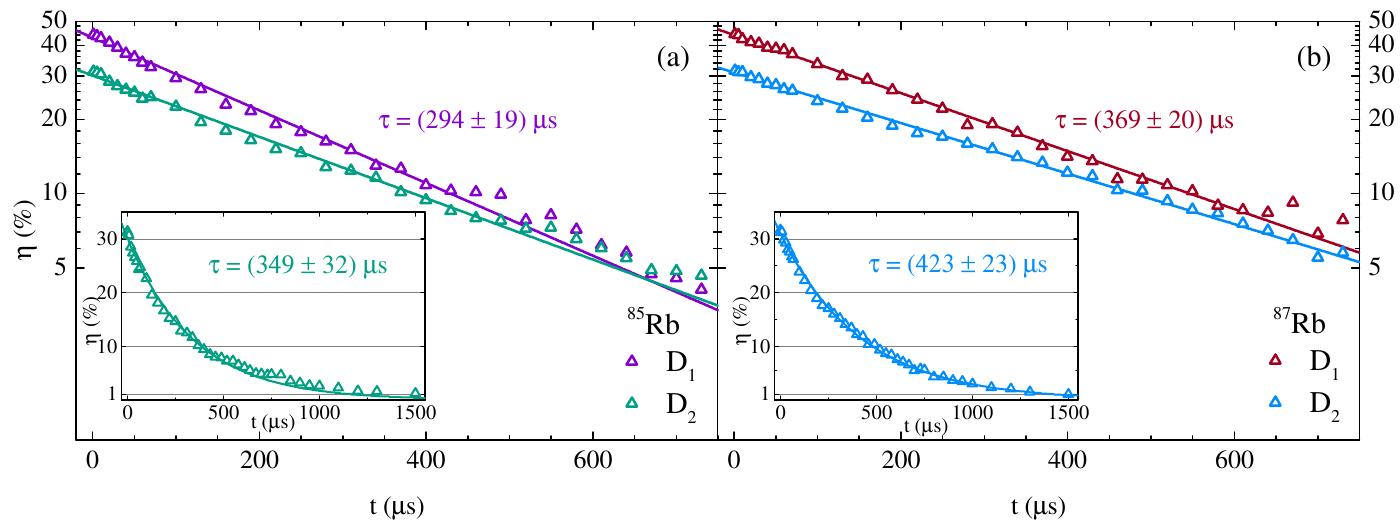}
\caption{\label{fig:lifetimes 85rb} Measured memory efficiency $\eta$ as a function of storage time for a vapor of $^{85}$Rb (a) and $^{87}$Rb (b) operating on the D$_1$ (violet, red) and D$_2$ (green, blue) atomic transitions and corresponding fits to the data. Measurements were performed at the cell temperature of $60^{\circ}$C, corresponding to the optical depth of OD = 9.5. The one- and two-photon detunings were set to values for which the highest measured efficiencies in Fig. \ref{fig:one_photon_scans 85rb} were obtained (i.e. operating points A,B,C and D). This corresponded to $\Delta = 900$ MHz and $\delta = 30$ kHz for the D$_1$ and $\Delta = -880$ MHz and $\delta = -20$ kHz for the D$_2$ of $^{85}$Rb and $\Delta = 900$ MHz and $\delta = 10$ kHz for the D$_1$ and $\Delta = -1000$ MHz and $\delta = -10$ kHz for the D$_2$ of $^{87}$Rb. Power of the coupling and the probe beams were adjusted so that the Rabi frequencies corresponded to $\Omega_{C} = 2 \pi \times 5.7$ MHz and $\Omega_{P} = 2 \pi \times 0.2$ MHz, respectively. Insets show measured measured memory efficiency $\eta$ as a function of storage time on $D_2$ transitions of the corresponding isotope, for longer storage times up to 1.5 ms.  }
\end{figure*}
\end{widetext}

\textbf{Memory lifetime.} In Fig. \ref{fig:lifetimes 85rb}, we show the measured memory efficiency $\eta$ (symbols) as a function of storage time for $^{85}\mathrm{Rb}$ (a) and $^{87}\mathrm{Rb}$ (b) vapors.
The measurements were performed at a cell temperature of $60^{\circ}$C using the D$_1$ (violet and red) and D$_2$ (green and blue) transitions under one-photon and two-photon detuning conditions that yielded the highest measured efficiencies. 
These operating points are labeled A, C, B, and D in Fig.~\ref{fig:one_photon_scans 85rb}, respectively.
All other experimental parameters are the same as those used in the measurements shown in Fig. \ref{fig:one_photon_scans 85rb} (a) and (b). The solid lines represent fits of the exponential function $\eta = A e^{-t/\tau}$ to the measured data, where $\tau$ denotes the memory lifetime. 

From the fits, we extract memory lifetimes of: (a) $\tau_{85}^{D_{1}} = 294 \pm 19~\mu$s at $\Delta = 900$~MHz and $\delta = +30$~kHz, and $\tau_{85}^{D_{2}} = 349 \pm 32~\mu$s at $\Delta = -880$~MHz and $\delta = -20$~kHz, for the D$_1$ and D$_2$ transitions of the $^{85}$Rb isotope, respectively and (b) $\tau_{87}^{D_{1}} = 369 \pm 20$ $\mu$s at $\Delta = 900$ MHz and $\delta = 10$ kHz and $\tau_{87}^{D_{2}} = 423 \pm 23$ $\mu$s at $\Delta = -1000$ MHz and $\delta = -10$ kHz, on the D$_1$ and D$_2$ transitions of the $^{87}$Rb isotope, respectively. The uncertainties of lifetimes were calculated by adding in quadrature the fit uncertainty and the statistical uncertainty from independent measurements of memory efficiency.


The memory lifetime is limited by the ground-state decoherence rate $\gamma_{12}$ averaged over the entire atomic ensemble.
At low optical depth, decoherence is mainly determined by: atomic motion causing finite interaction time of atoms with the optical fields, atom-atom collisions, residual Doppler broadening, magnetic field inhomogeneities and relative phase fluctuations of the probe and coupling laser fields.
At high optical depth, other processes can become dominant and shorten the ground-state coherence lifetime, such as radiation trapping, spontaneous Raman scattering and four wave mixing \cite{Novikova2011}.

In our case, the effects of relative phase fluctuations on decoherence can be neglected, since the coupling and probe beams are derived from the same laser source. 
Magnetic field inhomogeneities were estimated to be below $5\,\mathrm{mG}$, which corresponds to the measurement accuracy of the sensor used to characterize the magnetic field inside the three-layer $\mu$-metal shield.
The characteristic times for inelastic (spin-flip) Rb-Rb collisions are approximately 5.2~ms for $^{85}\mathrm{Rb}$ and 5.5~ms for $^{87}\mathrm{Rb}$. These values are calculated using
$\tau_{\mathrm{sf}} = 1/(N_{\mathrm{at}}\,\sigma_{\mathrm{Rb-Rb}}\,v_{\mathrm{Rb-Rb}})$,
where $N_{\mathrm{at}} = 2.45 \times 10^{11}~\mathrm{cm}^{-3}$ is the rubidium atomic number density at $60^\circ\mathrm{C}$, $\sigma_{\mathrm{Rb-Rb}}$ is the spin-exchange cross section taken from Ref. \cite{Ressler1969}, and $v_{\mathrm{Rb-Rb}}$ is the relative velocity between rubidium atoms.
Rb-Rb and Rb-Ne elastic collisions do not contribute to the dephasing of the ground states.


Furthermore, decoherence due to finite interaction times as well as residual Doppler broadening are strongly suppressed by the presence of 5~Torr of Ne buffer gas in the cells. 
Due to thermal motion, atoms interact with the optical fields of the probe and coupling beams only for a finite time. Upon leaving the interaction region, atoms lose spin-state coherence through collisions with the cell walls, representing the dominant decoherence mechanism associated with finite interaction times. 
In cells containing a noble buffer gas, frequent velocity-changing collisions induce diffusive atomic motion, thereby increasing the effective transit time through the interaction region and suppressing decoherence.
For our experimental conditions, we estimate a difussion transit time of approximately \(1.9\,\mathrm{ms}\), while the difference in diffusion time for the two isotopes is negligible.
The calculation is based on Eq.~(2) from Ref. \cite{Arimondo1996}
, taking into account 5~Torr of Ne buffer gas, and a probe beam radius of $R = 5.5$~mm. 
The diffusion coefficient, \(D = 27~\mathrm{cm}^2/\mathrm{s}\), is obtained using Eq.~(V.117) from Ref.~\cite{Happer1972}
, with the parameter \(D_0\) taken from \cite{Parniak2013}.
The mean free path is similar for both isotopes and is estimated to be approximately \(4.9~\mu\mathrm{m}\), based on the relations reported in \cite{Shuker2007}.
Due to its larger mass, the mean free path of the \(^{87}\mathrm{Rb}\) isotope differs by approximately \(1\%\) from that of \(^{85}\mathrm{Rb}\).
For comparison, in the absence of buffer gas, the transit time of \(^{87}\mathrm{Rb}\) atoms is approximately \(49~\mu\mathrm{s}\), as estimated from the probe beam diameter and the average two-dimensional atomic velocity. This is approximately 40 times shorter than the calculated transit time in the presence of 5~Torr of buffer gas.

For the co-propagating geometry used in our experiment, where the angle between the coupling and probe wavevectors $\mathbf{k}_c$ and $\mathbf{k}_p$ is close to zero, the decoherence associated with residual Doppler broadening in the Dicke regime scales as $D|\Delta \mathbf{k}|^2$, where $\Delta \mathbf{k} = \mathbf{k}_p - \mathbf{k}_c$ \cite{Firstenberg2008}. 
Using \(D = 27\,\mathrm{cm}^2/\mathrm{s}\), $|\Delta \mathbf{k}|= 63~\mathrm{m}^{-1}$ for $^{85}$Rb and $|\Delta \mathbf{k}|=143~\mathrm{m}^{-1}$ for $^{87}$Rb, we obtain characteristic Dicke-limited dephasing times of approximately 90~ms and 17.8~ms, respectively.

From the discussion above, the shortest decoherence timescale of 1.9~ms is associated with diffusion of atoms out of the interaction region of the laser fields and is the same for both isotopes. 
However, measurements of the memory efficiency as a function of storage time, shown in Fig.~\ref{fig:lifetimes 85rb} (a) and (b), indicate an approximately five times shorter memory lifetime.
This additional decoherence may arise from several sources, including magnetic-field inhomogeneities across the cell volume.
In our experiment, all magnetic sub-levels participate in the formation of the ground-state coherence, as no optical pumping into a specific \(m_F\) component is performed prior to the memory protocol. 
Weak stray magnetic fields that are insufficiently attenuated by the magnetic shielding can lift the degeneracy of the magnetic sub-levels. 
Furthermore, in the presence of magnetic-field non-uniformities, the energies of the magnetic sub-levels become position dependent. 
Due to the Brownian motion of atoms in the vapor, atoms migrating between regions of different magnetic-field strengths experience variations in Zeeman shifts, leading to effective redistribution among magnetic sub-levels.
This process results in unwanted broadening and contributes to the decoherence of the ground states \cite{DeRose2023}.

For example, in \(^{87}\mathrm{Rb}\), the \(F = 1\) ground state consists of three magnetic sublevels whose degeneracy is lifted in the presence of a weak external magnetic field through the Zeeman shift, \(\Delta E = g_F \mu_B m_F B_z\), where \(g_F\) is the Landé \(g\)-factor for the hyperfine state, \(\mu_B\) is the Bohr magneton, and \(m_F\) is the magnetic quantum number. 
The corresponding Zeeman splitting between adjacent magnetic sublevels in the \(F = 1\) state is approximately \(700~\mathrm{Hz/mG}\) \cite{Steck2003Rubidium87}
, implying that a magnetic field of \(1~\mathrm{mG}\) produces an energy separation of about \(1.4~\mathrm{kHz}\) between the \(m_F=\pm1\) states. 
This corresponds to a characteristic decoherence time of approximately \(114~\mu\mathrm{s}\). 
To reproduce a decoherence time comparable to the experimentally measured memory lifetime in \(^{87}\mathrm{Rb}\), magnetic-field inhomogeneities across the entire vapor-cell volume would need to be limited to approximately \(0.25~\mathrm{mG}\), corresponding to a characteristic decoherence time of about \(455~\mu\mathrm{s}\).
In the case of \(^{85}\mathrm{Rb}\), the lower \(F = 2\) ground state splits into five Zeeman sublevels. 
The Zeeman splitting between adjacent magnetic sublevels is approximately \(470~\mathrm{Hz/mG}\), such that a magnetic-field inhomogeneity of \(0.25~\mathrm{mG}\) would produce a maximum energy separation of about \(470~\mathrm{Hz}\) between the \(m_F=\pm2\) states. 
This corresponds to a characteristic decoherence time of approximately \(339~\mu\mathrm{s}\), consistent with a shorter observed memory lifetime compared to \(^{87}\mathrm{Rb}\).

Moreover, given the dense rubidium vapor conditions in our experiment (\(\mathrm{OD}=9.5\), \(N_{\mathrm{at}} = 2.45 \times 10^{11}~\mathrm{cm}^{-3}\)), radiation trapping, four-wave mixing (FWM), and Raman-scattered photons may additionally contribute to ground-state decoherence.
Although a detailed investigation of the influence of these effects on the memory lifetime is beyond the scope of this work, the existing literature suggests that they would be expected to favor the \(^{87}\mathrm{Rb}\) isotope in terms of achieving longer memory lifetimes, as observed in our work.
For example, the authors in \cite{Geng2014} 
investigated EIT in a cold atomic ensemble with large optical depth and reported a negligible FWM contribution in \(^{87}\mathrm{Rb}\), whereas a pronounced FWM signal was observed in the EIT output of \(^{85}\mathrm{Rb}\). 
This finding suggests that FWM-induced decoherence is expected to be stronger in \(^{85}\mathrm{Rb}\), leading to a faster loss of coherence.
In Ref.~\cite{DeRose2023}
, the authors discuss that a pump or coupling laser tuned near resonance to a specific \(F_g \rightarrow F_e\) transition can also off-resonantly excite nearby excited-state hyperfine levels, thereby introducing additional channels for atomic population leakage and reducing the steady-state population in the coherent state. 
They further conclude that the \(^{87}\mathrm{Rb}\) D$_1$ transition is advantageous in this regard, as its excited-state hyperfine splitting of \(816~\mathrm{MHz}\) is the largest, compared to \(362~\mathrm{MHz}\) for \(^{85}\mathrm{Rb}\) and the smaller splittings characteristic of the D2 transitions, leading to reduced off-resonant coupling.
And finally, it could be expected that radiation trapping becomes more detrimental in \(^{85}\mathrm{Rb}\) under the same experimental conditions due to its more complex hyperfine structure, which leads to enhanced reabsorption pathways and less efficient optical pumping compared to \(^{87}\mathrm{Rb}\).

The longest storage time we measured was 1.5 ms on the D$_2$ transition of $^{87}$Rb, with the memory efficiency of $1\%$. 


\section{\label{sec:conclusion}Conclusion}

To conclude, we investigated the optimal conditions for the realization of optical memory in warm rubidium vapor. Under the same experimental conditions, we characterized and compared the memory efficiencies and corresponding storage times for $^{85}$Rb and $^{87}$Rb isotopes. 
In addition, for each isotope we studied the efficiency and storage time using different hyperfine transitions on the D$_1$ and D$_2$ resonances. 
The experiment was performed using a $\Lambda$-type configuration consisting of two ground states and one excited state, with a strong coupling and a weak probe field, using a near-resonant EIT protocol.

It was shown that, for each transition and each isotope, exists an optimal one-photon detuning at which the memory efficiency is maximized. At this detuning, optimal conditions for light storage are achieved as a result of a trade-off between two competing effects: the reduction of coupling laser intensity due to absorption and the increased number of atoms within the relevant velocity class. 
For each one-photon detuning, the efficiency also depends on the two-photon detuning, therefore, it's optimization is crucial to obtain maximum memory efficiency.
Additionally, we have shown that the maximum efficiency increases with atomic vapor temperature, in the range of up to 60°C, limited by the paraffin coating of the cell. For each temperature of the cell, the "sweet spot" at which this maximum is obtained, should be searched for. Higher temperatures shift it toward larger absolute values of one-photon detuning.  

For both isotopes, higher maximum efficiencies were obtained on the D$_1$ transitions. 
When comparing the two isotopes, the highest achieved efficiencies of $44\%$ agree within $1\sigma$, while $^{87}$Rb exhibits longer memory lifetimes exceeding $400\,\mu$s.

Demonstrations of light storage in warm alkali vapors using EIT protocols typically optimize either efficiency (up to $\sim67 
\%$), at the expense of storage time (on the order of hundreds of nanoseconds) \cite{Ma2022}, 
or storage time (up to $1\,\mathrm{s}$), at the expense of efficiency ($\sim 10\%$) \cite{Katz2018}. 
In contrast, our work achieves a simultaneous optimization of both figures of merit by 
optimizing the one- and two-photon detunings of the laser beams. 
A similar optimization is expected to improve the performance of quantum memories in warm alkali vapors, particularly when combined with additional noise-suppression techniques developed in \cite{Namazi2017}.
This could pave the way for practical applications of quantum memories, including quantum device synchronization, buffering, and proof-of-principle quantum communication experiments, where both high efficiency and long storage times are essential.

\begin{acknowledgments}
This work was supported by the project Croatian Quantum Communication Infrastructure (CroQCI), funded by the European Union under the Digital Europe Programme (Grant No. 101091513) and by the Ministry of Science, Education and Youth of the Republic of Croatia; and by the project Quantum Enhanced Photonic Integrated Sensors For Metrology (QUANTIFY), which received funding from the European Union’s Horizon Europe research and innovation programme under Grant Agreement No. 101135931. The authors also acknowledge the project Centre for Advanced Laser Techniques (CALT), co-funded by the European Union through the European Regional Development Fund within the Competitiveness and Cohesion Operational Programme (Grant No. KK.01.1.1.05.0001). N.Š., D.A., and T.B. acknowledge support from the project "Implementation of cutting-edge research and its application as part of the Scientific Center of Excellence for Quantum and Complex Systems, and Representations of Lie Algebras", Grant No. PK.1.1.10.0004, co-financed by the European Union through the European Regional Development Fund -Competitiveness and Cohesion Programme 2021-2027. 
\end{acknowledgments}

\section*{Data Availability}
The data and code supporting the findings of this article are available upon reasonable request from the authors.

\bibliography{main_bib}

\end{document}